\documentclass[3p,times]{elsarticle}

\usepackage{ecrc}

\usepackage{epstopdf}

\volume{00}

\firstpage{1}

\journalname{Nuclear Physics A}

\runauth{B. Schenke and R.Venugopalan}

\jid{nupha}

\jnltitlelogo{Nuclear Physics A}

\usepackage{graphicx}
\usepackage{amsmath,amssymb}
\begin{document}

\begin{frontmatter}



\title{Collective effects in light-heavy ion collisions}

%% Single author (and collaboration) - please insert
%\author{Author (for the XYZ\fnref{col1} Collaboration)}
%\fntext[col1] {A list of members of the XYZ Collaboration and acknowledgements can be found at the end of this issue.}
%\address{Address}

%% For multiple authors, replace the above by:

\author{Bj\"orn Schenke and Raju Venugopalan}
%\author[label2]{Prithwish Tribedy}
%\author[]{}

\address{Physics Department, Brookhaven National Laboratory, Upton, NY 11973, USA}
%\address[label2]{Variable Energy Cyclotron Centre, 1/AF Bidhan Nagar, Kolkata 700064, India}

\begin{abstract}
We present results for the azimuthal anisotropy of charged hadron distributions in A+A, p+A, d+A, and $^3$He+A collisions
within the IP-Glasma+\textsc{music} model. Obtained  anisotropies are due to the fluid dynamic response
of the system to the fluctuating initial geometry of the interaction region. While the elliptic and triangular anisotropies in peripheral Pb+Pb collisions 
at $\sqrt{s}=2.76\,{\rm TeV}$ are well described by the model, the same quantities in $\sqrt{s}=5.02\,{\rm TeV}$ p+Pb collisions underestimate the experimental data. This disagreement can be due to neglected initial state correlations or the lack of a detailed description of the fluctuating
spatial structure of the proton, or both. We further present predictions for azimuthal anisotropies in p+Au, d+Au, and $^3$He+Au collisions at $\sqrt{s}=200\,{\rm GeV}$. For d+Au and $^3$He+Au collisions we expect the detailed substructure of the nucleon to become less important.
\end{abstract}

\begin{keyword}
heavy ion collisions \sep fluctuations \sep fluid dynamics

\end{keyword}

\end{frontmatter}

% \linenumbers


\section{Introduction}
\label{intro}
For the physical interpretation of observables in heavy ion, proton/deuteron-heavy ion and (high-multiplicity) proton+proton collisions, 
the existence of a sophisticated description of the multi-particle production mechanism and event-by-event fluctuations is essential.
In the high energy limit the color glass condensate (CGC) framework \cite{Gelis:2010nm} is the proper effective theory of quantum chromodynamics (QCD) 
that provides such description. 

One particular implementation of the CGC is the IP-Glasma model \cite{Schenke:2012wb,Schenke:2012hg}. It combines the IP-Sat dipole model \cite{Kowalski:2003hm,Kowalski:2007rw}, which parametrizes the impact parameter and $x$-dependence of the saturation scale, with the classical dynamics of produced gluon fields \cite{Krasnitz:1999wc,Krasnitz:2000gz,Lappi:2003bi}. %Bartels:2002cj,
With its parameters constrained by inclusive and diffractive deeply inelastic scattering (DIS) data from e+p scattering at HERA, the IP-Glasma model correctly describes the bulk features of various collision systems over a wide range of energies \cite{Schenke:2013dpa}. 

After briefly introducing the IP-Glasma model and the employed relativistic fluid dynamic simulation \textsc{music}, we 
present results for the azimuthal anisotropy of charged hadrons produced in peripheral $\sqrt{s}=2.76\,{\rm TeV}$ Pb+Pb collisions, and $\sqrt{s}=5.02\,{\rm TeV}$ p+Pb collisions of comparable multiplicity. We compare to experimental data from the CMS collaboration \cite{Chatrchyan:2013nka}.
We then present predictions for the systematics of the transverse momentum dependent anisotropy coefficients $v_n(p_T)$ in p+Au, d+Au, and $^3$He+Au collisions at top RHIC energies.

\section{IP-Glasma + MUSIC}\label{sec:ipglasma}
The IP-Glasma model \cite{Schenke:2012wb,Schenke:2012hg} relates the DIS constrained nuclear dipole cross-sections to the initial classical dynamics of highly occupied gluon fields produced in a nuclear collision. Given an initial distribution of color charges in the high energy nuclear wave-functions, the strong multiple scatterings of gluon fields are computed by event-by-event solutions of Yang-Mills equations. 
Both fluctuating distributions of nucleons in the nuclear wave-functions and intrinsic fluctuations of the color charge distributions are included. This results in ``lumpy'' transverse projections of the gluon field configurations that vary event to event. The scale of this lumpiness is given on average by the nuclear saturation scale $Q_s$ which corresponds to distance scales smaller than the nucleon size \cite{Kowalski:2007rw}.

%A detailed description of the IP-Glasma model can be found in \cite{Schenke:2012wb,Schenke:2012hg,Schenke:2013dpa}. 
%For the results shown in this work, the same model parameters as in \cite{Schenke:2013dpa} were used.

The IP-Glasma model provides the initial conditions for fluid dynamic calculations at a given time $\tau_0$. The initial energy density $\varepsilon$ and flow velocities $u^\mu$, are extracted from the gluon fields' energy-momentum tensor $T^{\mu\nu}$ at every transverse position via the relation $u_\mu T^{\mu\nu} = \varepsilon u^\nu$. 
In the results presented below, the viscous part of the energy momentum tensor is set to zero at the initial time of the fluid dynamic simulation. This is done because the gluon field strength tensor $T^{\mu\nu}$ is very anisotropic (the longitudinal pressure is approximately zero). A full 3+1 dimensional simulation including quantum fluctuations could provide a mechanism for isotropization via instabilities \cite{Berges:2013eia,Gelis:2013rba}.

We employ the viscous relativistic fluid dynamic simulation \textsc{music} \cite{Schenke:2010nt,Schenke:2010rr,Schenke:2011bn}, which is a 3+1 dimensional simulation. However, because the initial conditions from the IP-Glasma model are boost-invariant, it is used in its 2+1 dimensional mode. 

\section{p+Pb collisions at the LHC}
\begin{figure}[th]
\begin{center}
\includegraphics[width=0.48\textwidth]{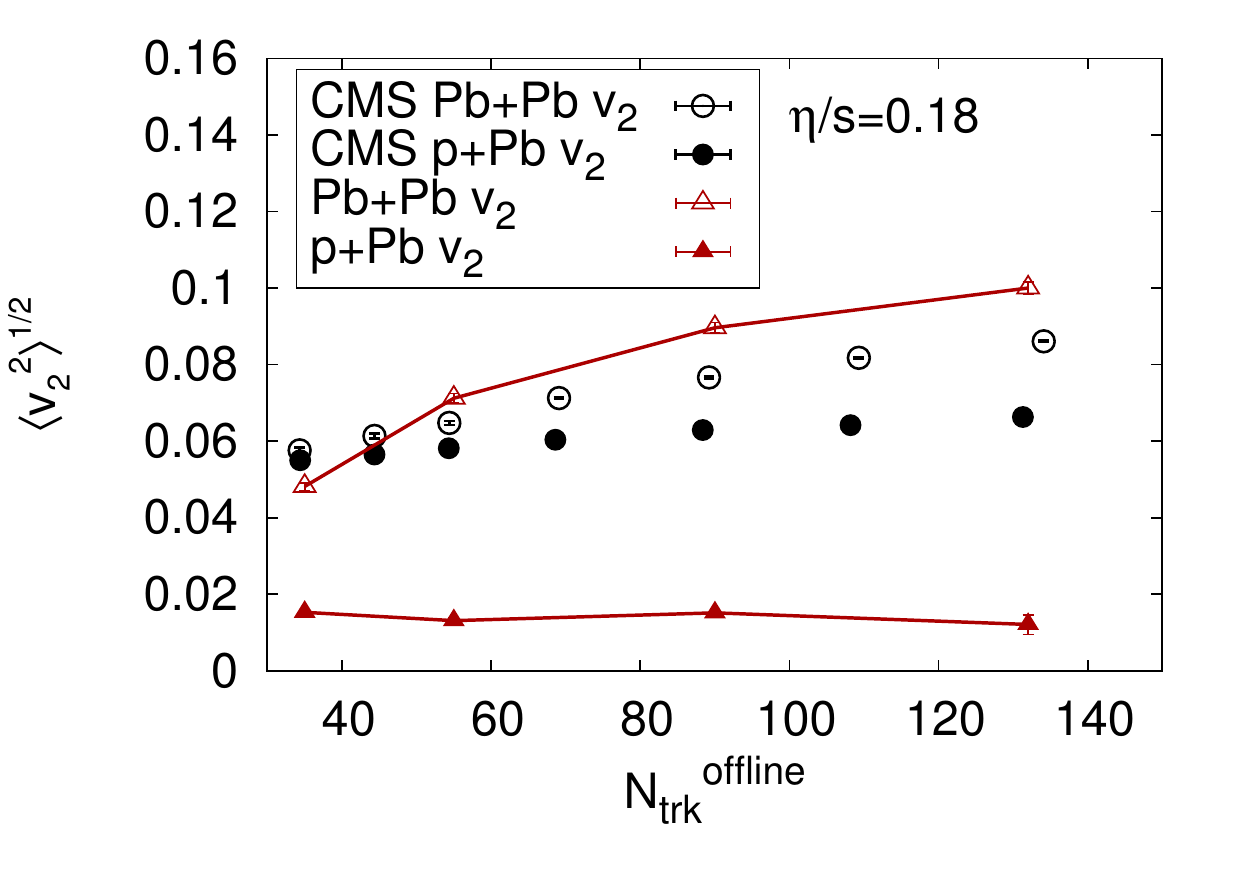}~\includegraphics[width=0.48\textwidth]{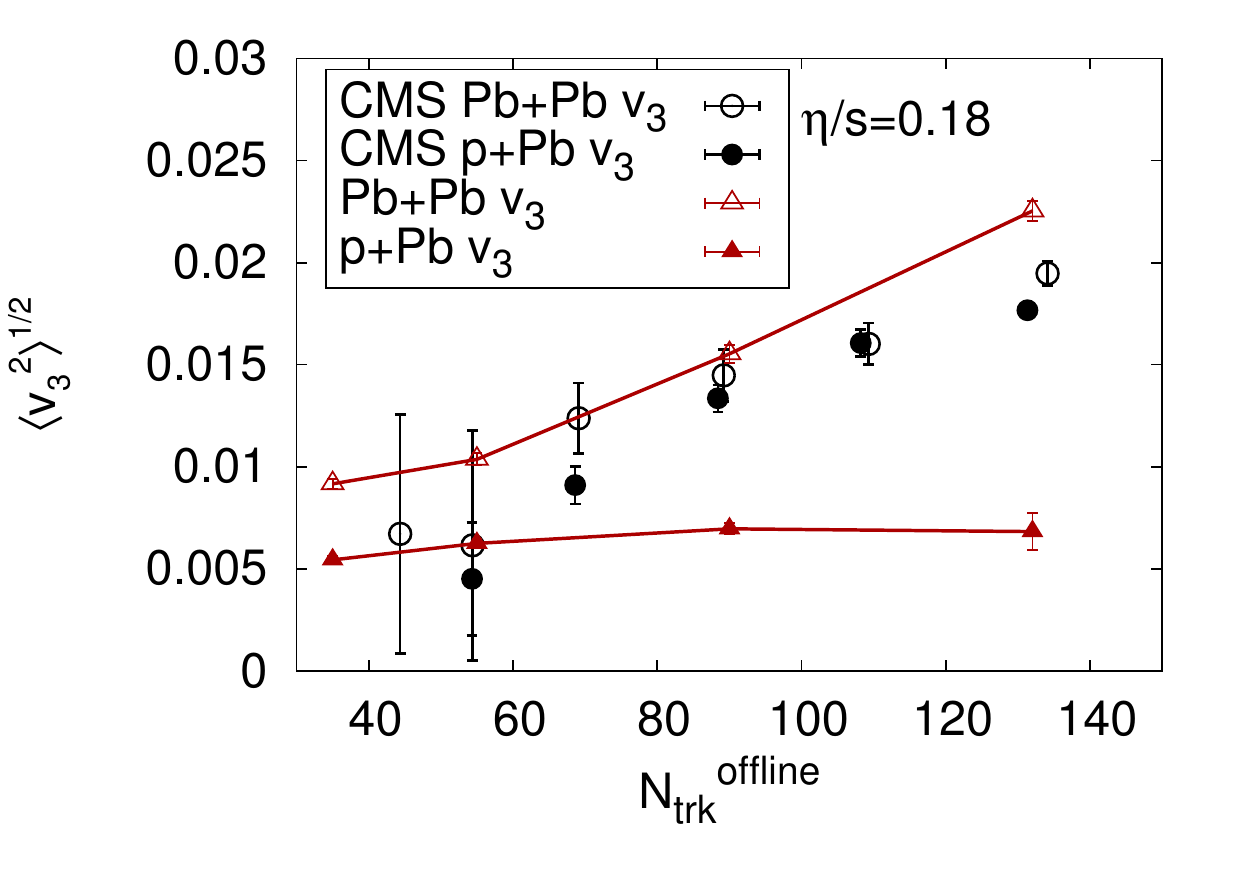}\\
\caption{Multiplicity dependence of the root-mean-square elliptic (left) and triangular (right) flow coefficients in Pb+Pb (open symbols) and p+Pb collisions (filled symbols) from the IP-Glasma+\textsc{music}
model (connected triangles) compared to experimental data by the CMS collaboration \cite{Chatrchyan:2013nka}. Figure from \cite{Schenke:2014zha}. \label{fig:vnCent-pPb-CMS}}\vspace{-0.5cm}
\end{center}
\end{figure}
It was shown in \cite{Schenke:2014zha} that experimental data from heavy ion collisions is well described by the IP-Glasma+\textsc{music} model out to fairly peripheral centrality bins. The natural question that arises is how the model does in describing data from p+A collisions that produce similar multiplicities.
Fig.\,\ref{fig:vnCent-pPb-CMS} shows the multiplicity dependence of $v_2$ and $v_3$ in peripheral Pb+Pb collisions and p+Pb collisions with comparable multiplicity. While the agreement with experimental data from the CMS collaboration \cite{Chatrchyan:2013nka} is fairly good in the Pb+Pb
case, for p+Pb collisions $v_2$ is under-predicted by approximately a factor of 4. In p+Pb $v_3$ agrees for the lower multiplicities studied, but has a rather flat multiplicity dependence and underestimates the experimental data at the higher multiplicities.

One reason for the disagreement could be that all initial state correlations that lead to an elliptic anisotropy\footnote{When including fluctuations initial state correlations should also contribute to odd harmonics.} \cite{Dusling:2012iga,Dusling:2012cg,Dusling:2012wy,Dusling:2013oia,Gyulassy:2014cfa,Dumitru:2014dra,McLerran:2014uka} are neglected. 
Additionally, the description of the proton in the IP-Glasma model is over-simplified. The proton shape is approximated by a sphere, and any deviations from that are due to the small scale color charge fluctuations. Since the interaction region in p+Pb collisions is dominated by the shape of the smaller projectile, initial geometries in p+Pb collisions have very small eccentricities. This inevitably leads to small $v_n$ coefficients.
If there is indeed a large contribution to $v_n$ coefficients in p+Pb collisions from collective effects, our result indicates that the shape of the proton fluctuates significantly more than assumed in the IP-Glasma model. One could envision a description where the small $x$ gluon distributions are still concentrated around large $x$ valence quark positions, leading to much larger eccentricities and fluctuations.
In this case p+A collisions could be used to determine the shape and fluctuations of gluon distributions in the proton at high energies. 

\section{p+Au, d+Au, and $^3$He+Au collisions at RHIC}
To determine whether final state collective effects provide the dominant contribution to the measured azimuthal anisotropy, RHIC is now
studying $^3$He+Au collisions that on average generate more triangular initial state configurations compared to p+Au or d+Au.
If collectivity is the physical explanation for the observed anisotropies, we expect a larger $v_3$ in $^3$He+Au collisions compared to p+Au and d+Au collisions at the same energy. To make this expectation more quantitative, we present predictions from the IP-Glasma+\textsc{music} framework.

For deuteron-gold collisions (d+Au) we compute the nucleon distribution in the deuteron using the Hulthen form of its wave function \cite{Miller:2007ri,Bzdak:2013zma}. For $^3$He, we use the same nucleon configurations as employed in \cite{Nagle:2013lja}. They are obtained from Green's
function Monte Carlo calculations using the AV18 + UIX model interaction \cite{Carlson:1997qn}.

For this comparative study we do not perform a detailed centrality selection, but instead sample the impact parameter $b$ between 0 and $2\,{\rm fm}$ in all systems. We then compute the initial state distribution of the energy density and flow velocity at time $\tau_0=0.5\,{\rm fm}/c$ and evolve the system using viscous fluid dynamics with $\eta/s=0.12$ until freeze-out at $T=135\,{\rm MeV}$. 

\begin{figure}
\begin{center}
\includegraphics*[width=0.96\textwidth]{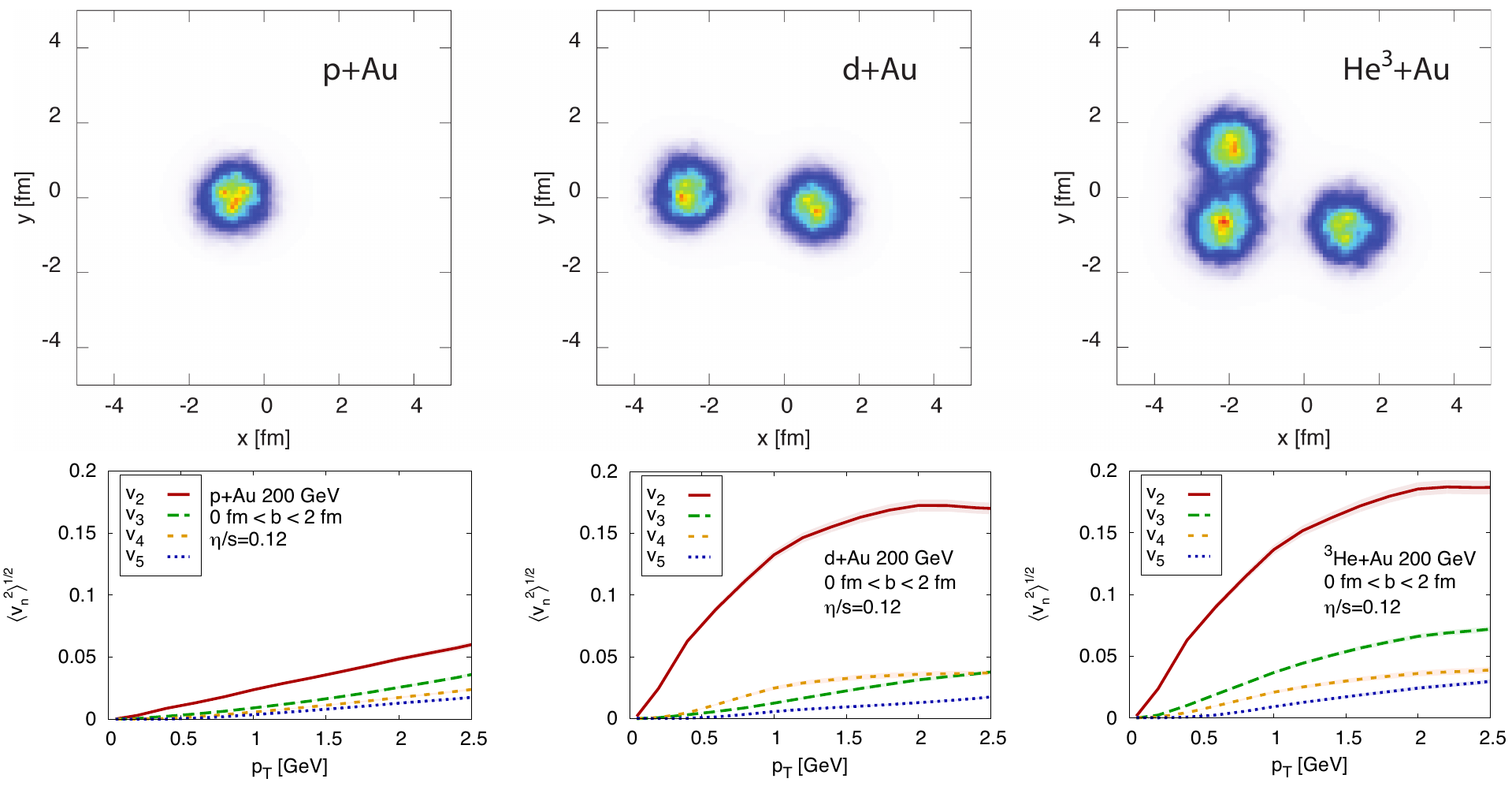}\\
\caption{Typical configurations of the initial energy density distribution for p+Au, d+Au and $^3$He+Au collisions (upper panel).
  Azimuthal anisotropy coefficients $v_2-v_5$ in the three collision systems (lower panel).\label{fig:pAu-dAu-He3Au-withFlow}}
\label{fig:generic}
\end{center}
\end{figure}

We present typical configurations of the initial energy density distribution in the transverse plane and final results for the transverse momentum dependent azimuthal anisotropy coefficients $v_2$ to $v_5$ in Fig.\,\ref{fig:pAu-dAu-He3Au-withFlow}.
While we find very small values for $v_2$ through $v_5$ in p+Au collisions, the additional nucleons and their position fluctuations generate larger $v_2-v_4$ in d+Au and $^3$He+Au collisions. The odd harmonics $v_3$ and $v_5$ are noticeably larger in $^3$He+Au collisions compared to d+Au collisions. This qualitative prediction can be compared to future measurements at RHIC.

\section{Conclusions}

%We have presented results from themodel for the azimuthal anisotropy of produced charged hadrons in small collision systems including p+Pb at LHC energies and p+Au, d+Au, and $^3$He+Au at RHIC energies.

We have demonstrated that experimental results for $v_2$ and $v_3$ in proton-heavy ion collisions at the LHC are not well described by the IP-Glasma+\textsc{music} model. Reasons for this could be the neglected initial state correlations and/or the lack of a detailed description of the fluctuating subnucleonic structure of the proton. Our results for p+A collisions differ significantly from those in \cite{Bozek:2012gr,Werner:2013ipa,Kozlov:2014fqa,Bzdak:2014dia}, suggesting that the details of the initial shape in small systems are of paramount importance.

We predict an increase of both $v_3$ and $v_5$ in $^3$He+Au collisions compared to d+Au collisions, while the even harmonics are comparable in both systems. The detailed substructure of the nucleon is expected to be less important for the initial state geometry in these collisions compared to p+A collisions.

\vspace{0.25cm}
\emph{Acknowledgments}
We thank Jamie Nagle for providing us with the $^3$He nucleon configurations. This work was supported under DOE Contract No. DE-AC02-98CH10886. BPS acknowledges support from a DOE Office of Science Early Career Award. This research used resources of the National Energy Research Scientific Computing Center, which is supported by the DOE Office of Science under Contract No. DE-AC02-05CH11231.



%% References


%\bibliographystyle{h-elsevier}
\bibliographystyle{elsarticle-num-names}
\bibliography{spires_pt}

\begin{thebibliography}{31}
\providecommand{\natexlab}[1]{#1}
\providecommand{\url}[1]{\texttt{#1}}
\providecommand{\urlprefix}{URL }
\expandafter\ifx\csname urlstyle\endcsname\relax
  \providecommand{\doi}[1]{doi:\discretionary{}{}{}#1}\else
  \providecommand{\doi}[1]{doi:\discretionary{}{}{}\begingroup
  \urlstyle{rm}\url{#1}\endgroup}\fi
\providecommand{\bibinfo}[2]{#2}

\bibitem[{Gelis et~al.(2010)Gelis, Iancu, Jalilian-Marian, and
  Venugopalan}]{Gelis:2010nm}
\bibinfo{author}{F.~Gelis}, \bibinfo{author}{E.~Iancu},
  \bibinfo{author}{J.~Jalilian-Marian}, \bibinfo{author}{R.~Venugopalan},
  \bibinfo{title}{{The Color Glass Condensate}},
  \bibinfo{journal}{Ann.Rev.Nucl.Part.Sci.} \bibinfo{volume}{60}
  (\bibinfo{year}{2010}) \bibinfo{pages}{463--489}.

\bibitem[{Schenke et~al.(2012{\natexlab{a}})Schenke, Tribedy, and
  Venugopalan}]{Schenke:2012wb}
\bibinfo{author}{B.~Schenke}, \bibinfo{author}{P.~Tribedy},
  \bibinfo{author}{R.~Venugopalan}, \bibinfo{title}{{Fluctuating Glasma initial
  conditions and flow in heavy ion collisions}}, \bibinfo{journal}{Phys. Rev.
  Lett.} \bibinfo{volume}{108} (\bibinfo{year}{2012}{\natexlab{a}})
  \bibinfo{pages}{252301}.

\bibitem[{Schenke et~al.(2012{\natexlab{b}})Schenke, Tribedy, and
  Venugopalan}]{Schenke:2012hg}
\bibinfo{author}{B.~Schenke}, \bibinfo{author}{P.~Tribedy},
  \bibinfo{author}{R.~Venugopalan}, \bibinfo{title}{{Event-by-event gluon
  multiplicity, energy density and eccentricities at RHIC and LHC}},
  \bibinfo{journal}{Phys. Rev.} \bibinfo{volume}{C86}
  (\bibinfo{year}{2012}{\natexlab{b}}) \bibinfo{pages}{034908}.

\bibitem[{Kowalski and Teaney(2003)}]{Kowalski:2003hm}
\bibinfo{author}{H.~Kowalski}, \bibinfo{author}{D.~Teaney}, \bibinfo{title}{{An
  impact parameter dipole saturation model}}, \bibinfo{journal}{Phys. Rev.}
  \bibinfo{volume}{D68} (\bibinfo{year}{2003}) \bibinfo{pages}{114005}.

\bibitem[{Kowalski et~al.(2008)Kowalski, Lappi, and
  Venugopalan}]{Kowalski:2007rw}
\bibinfo{author}{H.~Kowalski}, \bibinfo{author}{T.~Lappi},
  \bibinfo{author}{R.~Venugopalan}, \bibinfo{title}{{Nuclear enhancement of
  universal dynamics of high parton densities}}, \bibinfo{journal}{Phys. Rev.
  Lett.} \bibinfo{volume}{100} (\bibinfo{year}{2008}) \bibinfo{pages}{022303}.

\bibitem[{Krasnitz and Venugopalan(2000)}]{Krasnitz:1999wc}
\bibinfo{author}{A.~Krasnitz}, \bibinfo{author}{R.~Venugopalan},
  \bibinfo{title}{The initial energy density of gluons produced in very high
  energy nuclear collisions}, \bibinfo{journal}{Phys. Rev. Lett.}
  \bibinfo{volume}{84} (\bibinfo{year}{2000}) \bibinfo{pages}{4309--4312}.

\bibitem[{Krasnitz and Venugopalan(2001)}]{Krasnitz:2000gz}
\bibinfo{author}{A.~Krasnitz}, \bibinfo{author}{R.~Venugopalan},
  \bibinfo{title}{{The initial gluon multiplicity in heavy ion collisions}},
  \bibinfo{journal}{Phys. Rev. Lett.} \bibinfo{volume}{86}
  (\bibinfo{year}{2001}) \bibinfo{pages}{1717--1720}.

\bibitem[{Lappi(2003)}]{Lappi:2003bi}
\bibinfo{author}{T.~Lappi}, \bibinfo{title}{{Production of gluons in the
  classical field model for heavy ion collisions}}, \bibinfo{journal}{Phys.
  Rev.} \bibinfo{volume}{C67} (\bibinfo{year}{2003}) \bibinfo{pages}{054903}.

\bibitem[{Schenke et~al.(2014)Schenke, Tribedy, and
  Venugopalan}]{Schenke:2013dpa}
\bibinfo{author}{B.~Schenke}, \bibinfo{author}{P.~Tribedy},
  \bibinfo{author}{R.~Venugopalan}, \bibinfo{title}{{Multiplicity distributions
  in p+p, p+A and A+A collisions from Yang-Mills dynamics}},
  \bibinfo{journal}{Phys.Rev.} \bibinfo{volume}{C89} (\bibinfo{year}{2014})
  \bibinfo{pages}{024901}.

\bibitem[{Chatrchyan et~al.(2013)}]{Chatrchyan:2013nka}
\bibinfo{author}{S.~Chatrchyan}, et~al., \bibinfo{title}{{Multiplicity and
  transverse momentum dependence of two- and four-particle correlations in pPb
  and PbPb collisions}}, \bibinfo{journal}{Phys.Lett.} \bibinfo{volume}{B724}
  (\bibinfo{year}{2013}) \bibinfo{pages}{213--240}.

\bibitem[{Berges et~al.(2014)Berges, Boguslavski, Schlichting, and
  Venugopalan}]{Berges:2013eia}
\bibinfo{author}{J.~Berges}, \bibinfo{author}{K.~Boguslavski},
  \bibinfo{author}{S.~Schlichting}, \bibinfo{author}{R.~Venugopalan},
  \bibinfo{title}{{Turbulent thermalization process in heavy-ion collisions at
  ultrarelativistic energies}}, \bibinfo{journal}{Phys.Rev.}
  \bibinfo{volume}{D89} (\bibinfo{year}{2014}) \bibinfo{pages}{074011}.

\bibitem[{Epelbaum and Gelis(2013)}]{Gelis:2013rba}
\bibinfo{author}{T.~Epelbaum}, \bibinfo{author}{F.~Gelis},
  \bibinfo{title}{{Pressure isotropization in high energy heavy ion
  collisions}}, \bibinfo{journal}{Phys.Rev.Lett.} \bibinfo{volume}{111}
  (\bibinfo{year}{2013}) \bibinfo{pages}{232301}.

\bibitem[{Schenke et~al.(2010)Schenke, Jeon, and Gale}]{Schenke:2010nt}
\bibinfo{author}{B.~Schenke}, \bibinfo{author}{S.~Jeon},
  \bibinfo{author}{C.~Gale}, \bibinfo{title}{{(3+1)D hydrodynamic simulation of
  relativistic heavy-ion collisions}}, \bibinfo{journal}{Phys. Rev.}
  \bibinfo{volume}{C82} (\bibinfo{year}{2010}) \bibinfo{pages}{014903}.

\bibitem[{Schenke et~al.(2011{\natexlab{a}})Schenke, Jeon, and
  Gale}]{Schenke:2010rr}
\bibinfo{author}{B.~Schenke}, \bibinfo{author}{S.~Jeon},
  \bibinfo{author}{C.~Gale}, \bibinfo{title}{{Elliptic and triangular flow in
  event-by-event (3+1)D viscous hydrodynamics}}, \bibinfo{journal}{Phys. Rev.
  Lett.} \bibinfo{volume}{106} (\bibinfo{year}{2011}{\natexlab{a}})
  \bibinfo{pages}{042301}.

\bibitem[{Schenke et~al.(2011{\natexlab{b}})Schenke, Jeon, and
  Gale}]{Schenke:2011bn}
\bibinfo{author}{B.~Schenke}, \bibinfo{author}{S.~Jeon},
  \bibinfo{author}{C.~Gale}, \bibinfo{title}{{Higher flow harmonics from (3+1)D
  event-by-event viscous hydrodynamics}}, \bibinfo{journal}{Phys. Rev.}
  \bibinfo{volume}{C85} (\bibinfo{year}{2011}{\natexlab{b}})
  \bibinfo{pages}{024901}.

\bibitem[{Schenke and Venugopalan(2014)}]{Schenke:2014zha}
\bibinfo{author}{B.~Schenke}, \bibinfo{author}{R.~Venugopalan},
  \bibinfo{title}{{Eccentric protons? Sensitivity of flow to system size and
  shape in p+p, p+Pb and Pb+Pb collisions}} \bibinfo{note}{, arXiv:1405.3605}.

\bibitem[{Dusling and Venugopalan(2012)}]{Dusling:2012iga}
\bibinfo{author}{K.~Dusling}, \bibinfo{author}{R.~Venugopalan},
  \bibinfo{title}{{Azimuthal collimation of long range rapidity correlations by
  strong color fields in high multiplicity hadron-hadron collisions}},
  \bibinfo{journal}{Phys.Rev.Lett.} \bibinfo{volume}{108}
  (\bibinfo{year}{2012}) \bibinfo{pages}{262001}.

\bibitem[{Dusling and Venugopalan(2013{\natexlab{a}})}]{Dusling:2012cg}
\bibinfo{author}{K.~Dusling}, \bibinfo{author}{R.~Venugopalan},
  \bibinfo{title}{{Evidence for BFKL and saturation dynamics from dihadron
  spectra at the LHC}}, \bibinfo{journal}{Phys.Rev.}
  \bibinfo{volume}{D87}~(\bibinfo{number}{5})
  (\bibinfo{year}{2013}{\natexlab{a}}) \bibinfo{pages}{051502}.

\bibitem[{Dusling and Venugopalan(2013{\natexlab{b}})}]{Dusling:2012wy}
\bibinfo{author}{K.~Dusling}, \bibinfo{author}{R.~Venugopalan},
  \bibinfo{title}{{Explanation of systematics of CMS p+Pb high multiplicity
  di-hadron data at $\sqrt{s}_{\rm NN} = 5.02$ TeV}},
  \bibinfo{journal}{Phys.Rev.} \bibinfo{volume}{D87}~(\bibinfo{number}{5})
  (\bibinfo{year}{2013}{\natexlab{b}}) \bibinfo{pages}{054014}.

\bibitem[{Dusling and Venugopalan(2013{\natexlab{c}})}]{Dusling:2013oia}
\bibinfo{author}{K.~Dusling}, \bibinfo{author}{R.~Venugopalan},
  \bibinfo{title}{{Comparison of the Color Glass Condensate to di-hadron
  correlations in proton-proton and proton-nucleus collisions}},
  \bibinfo{journal}{Phys.Rev.} \bibinfo{volume}{D87}
  (\bibinfo{year}{2013}{\natexlab{c}}) \bibinfo{pages}{094034}.

\bibitem[{Gyulassy et~al.(2014)Gyulassy, Levai, Vitev, and
  Biro}]{Gyulassy:2014cfa}
\bibinfo{author}{M.~Gyulassy}, \bibinfo{author}{P.~Levai},
  \bibinfo{author}{I.~Vitev}, \bibinfo{author}{T.~Biro},
  \bibinfo{title}{{Non-Abelian Bremsstrahlung and Azimuthal Asymmetries in High
  Energy p+A Reactions}} \bibinfo{note}{, arXiv:1405.7825}.

\bibitem[{Dumitru and Giannini(2014)}]{Dumitru:2014dra}
\bibinfo{author}{A.~Dumitru}, \bibinfo{author}{A.~V. Giannini},
  \bibinfo{title}{{Initial state angular asymmetries in high energy p+A
  collisions: spontaneous breaking of rotational symmetry by a color electric
  field and C-odd fluctuations}} \bibinfo{note}{, arXiv:1406.5781}.

\bibitem[{McLerran and Skokov(2014)}]{McLerran:2014uka}
\bibinfo{author}{L.~McLerran}, \bibinfo{author}{V.~V. Skokov},
  \bibinfo{title}{{The Eccentric Collective BFKL Pomeron}} \bibinfo{note}{,
  arXiv:1407.2651}.

\bibitem[{Miller et~al.(2007)Miller, Reygers, Sanders, and
  Steinberg}]{Miller:2007ri}
\bibinfo{author}{M.~L. Miller}, \bibinfo{author}{K.~Reygers},
  \bibinfo{author}{S.~J. Sanders}, \bibinfo{author}{P.~Steinberg},
  \bibinfo{title}{{Glauber modeling in high energy nuclear collisions}},
  \bibinfo{journal}{Ann. Rev. Nucl. Part. Sci.} \bibinfo{volume}{57}
  (\bibinfo{year}{2007}) \bibinfo{pages}{205--243}.

\bibitem[{Bzdak et~al.(2013)Bzdak, Schenke, Tribedy, and
  Venugopalan}]{Bzdak:2013zma}
\bibinfo{author}{A.~Bzdak}, \bibinfo{author}{B.~Schenke},
  \bibinfo{author}{P.~Tribedy}, \bibinfo{author}{R.~Venugopalan},
  \bibinfo{title}{{Initial state geometry and the role of hydrodynamics in
  proton-proton, proton-nucleus and deuteron-nucleus collisions}},
  \bibinfo{journal}{Phys.Rev.} \bibinfo{volume}{C87}~(\bibinfo{number}{6})
  (\bibinfo{year}{2013}) \bibinfo{pages}{064906}.

\bibitem[{Nagle et~al.(2013)Nagle, Adare, Beckman, Koblesky, Koop
  et~al.}]{Nagle:2013lja}
\bibinfo{author}{J.~Nagle}, \bibinfo{author}{A.~Adare},
  \bibinfo{author}{S.~Beckman}, \bibinfo{author}{T.~Koblesky},
  \bibinfo{author}{J.~O. Koop}, et~al., \bibinfo{title}{{Exploiting Intrinsic
  Triangular Geometry in Relativistic He3+Au Collisions to Disentangle Medium
  Properties}} \bibinfo{note}{, arXiv:1312.4565}.

\bibitem[{Carlson and Schiavilla(1998)}]{Carlson:1997qn}
\bibinfo{author}{J.~Carlson}, \bibinfo{author}{R.~Schiavilla},
  \bibinfo{title}{{Structure and dynamics of few nucleon systems}},
  \bibinfo{journal}{Rev.Mod.Phys.} \bibinfo{volume}{70} (\bibinfo{year}{1998})
  \bibinfo{pages}{743--842}.

\bibitem[{Bozek and Broniowski(2013)}]{Bozek:2012gr}
\bibinfo{author}{P.~Bozek}, \bibinfo{author}{W.~Broniowski},
  \bibinfo{title}{{Correlations from hydrodynamic flow in p-Pb collisions}},
  \bibinfo{journal}{Phys.Lett.} \bibinfo{volume}{B718} (\bibinfo{year}{2013})
  \bibinfo{pages}{1557--1561},
  \doi{\bibinfo{doi}{10.1016/j.physletb.2012.12.051}}.

\bibitem[{Werner et~al.(2014)Werner, Bleicher, Guiot, Karpenko, and
  Pierog}]{Werner:2013ipa}
\bibinfo{author}{K.~Werner}, \bibinfo{author}{M.~Bleicher},
  \bibinfo{author}{B.~Guiot}, \bibinfo{author}{I.~Karpenko},
  \bibinfo{author}{T.~Pierog}, \bibinfo{title}{{Evidence for flow in pPb
  collisions at 5 TeV from v2 mass splitting}},
  \bibinfo{journal}{Phys.Rev.Lett.} \bibinfo{volume}{112}
  (\bibinfo{year}{2014}) \bibinfo{pages}{232301},
  \doi{\bibinfo{doi}{10.1103/PhysRevLett.112.232301}}.

\bibitem[{Kozlov et~al.(2014)Kozlov, Luzum, Denicol, Jeon, and
  Gale}]{Kozlov:2014fqa}
\bibinfo{author}{I.~Kozlov}, \bibinfo{author}{M.~Luzum},
  \bibinfo{author}{G.~Denicol}, \bibinfo{author}{S.~Jeon},
  \bibinfo{author}{C.~Gale}, \bibinfo{title}{{Transverse momentum structure of
  pair correlations as a signature of collective behavior in small collision
  systems}} \bibinfo{note}{, arXiv:1405.3976}.

\bibitem[{Bzdak and Ma(2014)}]{Bzdak:2014dia}
\bibinfo{author}{A.~Bzdak}, \bibinfo{author}{G.-L. Ma},
  \bibinfo{title}{{Elliptic and triangular flow in p+Pb and peripheral Pb+Pb
  collisions from parton scatterings}} \bibinfo{note}{, arXiv:1406.2804}.

\end{thebibliography}



%\begin{thebibliography}{00}


%\bibitem{ref1} J. van der Geer, J.A.J. Hanraads, R.A. Lupton, J. Sci. Commun. 163 (2000) 51Ð59. 
%\bibitem{ref2} W. Strunk Jr., E.B. White, The Elements of Style, third ed., Macmillan, New York, 1979. 

%\end{thebibliography}

\end{document}